\documentclass[11pt,twoside]{article}  
\usepackage{adassconf}

\begin{document}   

%
%
%

\paperID{D06}

%
%
%
%

\title{AstroCat/CVcat: A catalogue on Cataclysmic Variables based on a new framework for online interactive astronomical databases}
\titlemark{AstroCat/CVcat: A catalogue on Cataclysmic Variables}

%
%
%

\author{Fabian Euchner, Alexander Pollmer, Klaus Beuermann}
\affil{Universit\"ats-Sternwarte, Geismarlandstr.~11, D-37083 G\"ottingen, Germany}
\author{Boris T.\ G\"ansicke}
\affil{Department of Physics, University of Warwick, Coventry CV4~7AL, UK}
\author{Jens Kube}
\affil{Stiftung Alfred-Wegener-Institut f\"ur Polar- und Meeresforschung in der Helmholtz-Gemeinschaft, Koldewey-Station, 
N-9173 Ny-{\AA}lesund, Norway}

%
%

\contact{Fabian Euchner}
\email{feuchner@astro.physik.uni-goettingen.de}

%
%
%
%
%

\paindex{Euchner, F.}
\aindex{Pollmer, A.}
\aindex{G\"ansicke, B. T.} 
\aindex{Kube, J.} 
\aindex{Beuermann, K.} 

%
%

\authormark{Euchner, Pollmer, G\"ansicke, Kube \& Beuermann}

%
%



\begin{abstract}          
We report on the progress of the development of CVcat, an interactive 
catalogue on Cataclysmic Variables, which is the first application based on
AstroCat, a general framework for the installation and maintenance of
web-based interactive astronomical databases. 
Registered users can contribute 
directly to the catalogue content by adding new objects, object properties, 
literature references, and annotations. 
The scientific quality control of the 
catalogue is carried out by a distributed editorial team.
Searches in CVcat can be performed by object name, classification,
certain properties or property ranges, and coordinates. Search results 
can be retrieved in several output formats, including XML.
Old database states can be restored in order to ensure the citability
of the catalogue. 
Furthermore, CVcat is designed to serve as a repository for 
reduced data from publications.
Future prospects include the integration of AstroCat-based catalogues
in the international network of Virtual Observatories.
\end{abstract}

%
%

\section{AstroCat -- a new concept for astronomical catalogues}
\subsection{Motivation}
Traditionally, in astronomy the availability of online digital information is excellent with respect
to scientific publications (NASA's \emph{Astrophysics Data System}, \emph{arXiv.org} preprint
server) and raw observational data. With the development of the \emph{AstroCat} software
we intend to fill the gap between these two categories by enabling astronomers to set up 
\emph{interactive} astronomical catalogues for reduced and inferred data (Fig.~1). 
\begin{figure}
\plotone{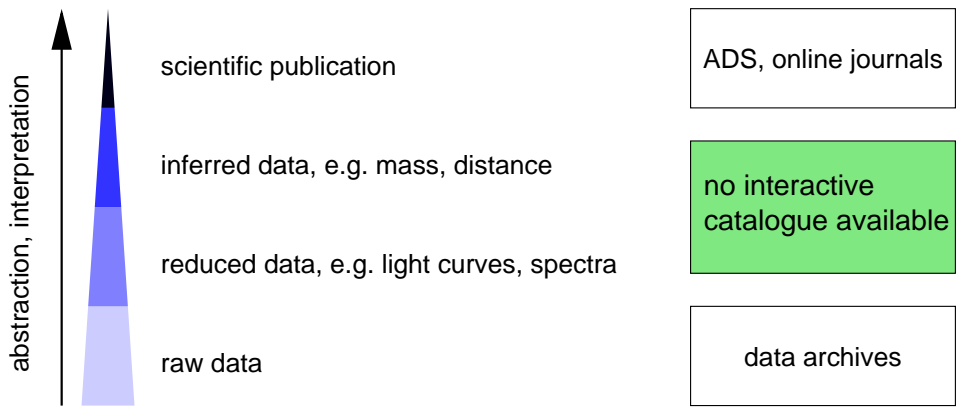}
\caption{Level of abstraction and digital availability of astronomical information.}
\end{figure}

\subsection{Objectives}
\emph{AstroCat} is a software framework for the implementation of a new type of web-based 
interactive astronomical catalogues.
\emph{AstroCat}-based catalogues are intended to hold information on the physical properties 
of astronomical objects of a specific class. We provide the possibility to account for 
fine-grained hierarchical sub-classifications of the selected objects. The information contained
in the catalogue should be gathered from trustworthy publications, preferably from refereed papers.
The scientific quality control is performed by users who assumed editorial duties. If available, 
a hyperlink to an electronic version of the referenced publication should be given for each catalogue 
entry (e.g.\ via the ADS, the astro-ph preprint server, or online journals like the 
\emph{Information Bulletin on Variable Stars}).
We will also provide the possibility for authors to share reduced data (spectra, light curves, images, etc.) from 
their publications with the community of users. 

Our new concept of astronomical catalogues and the differences to existing catalogues 
are best characterized by the terms \emph{interactivity}, \emph{up-to-dateness}
and \emph{accessibility}:

\emph{Interactivity:} All registered users may contribute to the database content by adding new data. 
The reliability of the data is ensured by an editorial team which is allowed to modify catalogue
entries. We achieve a high level of objectiveness by allowing for several entries per property. 
We also allow for detailed annotations on the catalogue entries.

\emph{Up-to-dateness:} Most `classical' catalogues are updated only in irregular and/or lengthy intervals. 
In \emph{AstroCat}-based catalogues all changes to the database are made instantly visible to the users. 
To ensure the citability of the catalogue, we provide a mechanism for restoring previous states of 
the catalogue content.

\emph{Accessibility:} The web-based character allows for simple but powerful searching on the 
database via a web browser. The query results can be formatted in various user-definable styles. 
We also provide the possibility to retrieve the query results in XML format in order to supply the user 
with semantically enriched data. 

Additional information on the \emph{AstroCat/CVcat} project can be found at our 
\htmladdnormallinkfoot{web page}{http://astrocat.uni-goettingen.de}.
At this location, we provide an online 
\htmladdnormallinkfoot{discussion forum}{http://astrocat.uni-goettingen.de/\#discussion} 
where comments on the project can be placed.

\section{CVcat -- the online catalogue on Cataclysmic Variables}
\emph{CVcat}, a first version of an online catalogue on Cataclysmic Variables (CVs), 
was developed by the CV group in G\"ottingen and presented to the public in August 2001 
due to the increasing need in the community of CV researchers for an authoritative, up-to-date, 
online database of the relevant objects (Kube et al. 2003). In this catalogue some of the concepts 
of \emph{AstroCat} are already realized. It comprises data from `classical' 
catalogues on CVs (mainly Ritter \& Kolb 2003) as well as additional information compiled 
manually from numerous publications.

Since the acceptance of \emph{CVcat} in the CV community is good, we decided to
re-implement the catalogue with additional features providing more flexibility and convenience 
to the users. For that purpose, we started to develop the \emph{AstroCat} framework which is not
only designed for the re-implementation of \emph{CVcat}, but can also be used for the 
installation of catalogues covering different astronomical fields.

Up to now, \emph{CVcat} is used by $\sim$150 registered users and can be accessed at 
\htmladdURL{http://www.cvcat.org}. The upgrade to the \emph{AstroCat}-based version is
planned for January 2004. A non-interactive demonstration of the new version can already be
found at \htmladdURL{http://astrocat.uni-goettingen.de/cvcat-demo/}. 

\section{Target Group}
\emph{AstroCat} is especially suited to set up databases used by relatively small research communities 
(several hundred users). Since all catalogue entries should be approved by editors, we estimate 
the maximum number of objects that can be handled properly to be several thousands. The possibility to
comment on catalogue entries is particularly useful if extensive calculations and/or non-standard
methods are required to derive the respective object properties. 

\section{Technical Realization}
The catalogue data is held in a PostgreSQL database management system. 
The communication between the database and the webserver (Apache) is controlled by PHP scripts. 
Queries to and results from the database are  handled internally in an XML dialect, 
\htmladdnormallinkfoot{\emph{AstroCatML}}{http://astrocat.uni-goettingen.de/\#astrocatml}, 
for which an XML Schema can be found on our web page.
A schematic view of the data flow in \emph{AstroCat} can be found in Fig.~2.
\begin{figure}
\plotone{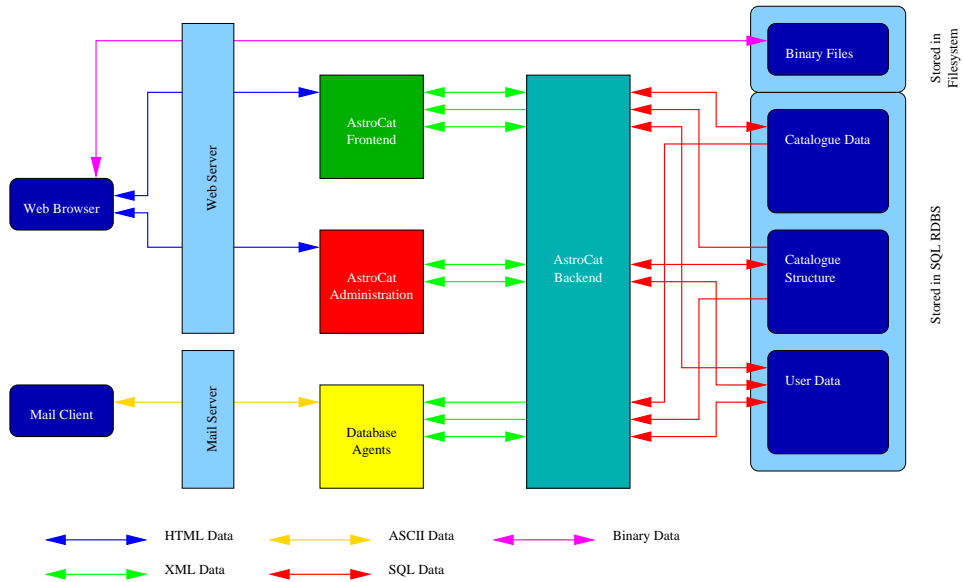}
\caption{Software components and data flow in AstroCat.}
\end{figure}

\section{Outlook}
After the completion of \emph{CVcat} we plan to transfer the new concept to a different field of
astronomical research by installing a catalogue on extrasolar planets \emph{(EPcat)} based on 
the \emph{AstroCat} software.
Furthermore, we will provide a registry where the metadata of all \emph{AstroCat}-based catalogues
can be stored, in order to establish interoperability mechanisms between catalogues, e.g.\ simultaneous searches.
It is also intended to integrate the information provided by the catalogues based on \emph{AstroCat}
in the global network of Virtual Observatories.

\section{Partners}
\emph{AstroCat/CVcat} is funded by the \emph{Deutsche Forschungsgemeinschaft}
(project number LIS 4 - 554 95 (1) SUB G\"ottingen). 
The project is realized in collaboration with the Nieders\"achsische Staats- und Universit\"atsbibliothek (SUB), 
G\"ottingen, in the framework of \emph{Virtuelle Fachbibliothek Astronomie}. 
\emph{CVcat} will be hosted at the SUB after completion.

\end{document}